\documentclass[preprint,showpacs,preprintnumbers,amsmath,amssymb,prb]{revtex4}

\usepackage{graphicx}
\usepackage{dcolumn}
\usepackage{bm}

\makeatletter
\def\@dotsep{4.5}
\makeatother

\begin{document}


\title{Second-order nonadiabatic couplings from time-dependent density functional 
theory:  Evaluation in the immediate vicinity of Jahn-Teller/Renner-Teller intersections    
}

\author{Chunping Hu$^1$}
\email{hu@rs.kagu.tus.ac.jp}
\author{Osamu Sugino$^2$}
\author{Kazuyuki Watanabe$^1$}
\affiliation{
$^1$Department of Physics,
Tokyo University of Science,
1-3 Kagurazaka, Shinjuku,
Tokyo 162-8601, Japan\\
$^2$Institute for Solid State Physics, University of Tokyo,
Kashiwa, Chiba 277-8581, Japan
} 

\date{\today}

\begin{abstract} 
For a rigorous quantum simulation of nonadiabatic dynamics of electrons and nuclei, 
knowledge of not only first-order but also second-order nonadiabatic couplings (NAC), 
is required. Here we propose a method to efficiently calculate second-order NAC 
from  time-dependent density functional theory (TDDFT), on the basis of the Casida ansatz adapted for the computation  
of first-order NAC, which has been justified in our previous work and can be shown to be valid for calculating  
second-order NAC between ground state and singly excited states within the Tamm-Dancoff approximation. Test calculations of 
second-order NAC in the immediate vicinity of Jahn-Teller and Renner-Teller intersections show that calculation 
results from TDDFT, combined with modified linear response theory, agree well with the prediction from 
the Jahn-Teller / Renner-Teller models. Contrary to the diverging behavior of first-order NAC 
near all types of intersection points, the Cartesian components of second-order NAC are shown to be 
negligibly small near Renner-Teller glancing intersections, while they are significantly large 
near the Jahn-Teller conical intersections. Nevertheless, the components of second-order NAC can cancel 
each other to a large extent in Jahn-Teller systems, indicating the background of neglecting 
second-order NAC in practical dynamics simulations.  On the other hand, it is shown that such a 
cancellation becomes less effective in an elliptic Jahn-Teller system and thus the role of 
second-order NAC needs to be evaluated in the rigorous framework.  Our study shows that TDDFT is 
promising to provide accurate data of NAC for full quantum mechanical simulation of nonadiabatic 
processes.    

\end{abstract}

\maketitle

\section{Introduction\label{intro}} 

Nonadiabatic transitions, i.e., transitions between adiabatic states, 
are ubiquitous in physical, chemical and biological systems. \cite{yarkony-RMP,nakamura-book,mbaer-book} 
In recent years there has been growing interest in quantum mechanical study of 
nonadiabatic transitions,\cite{marx-prl,xiaosong,li-tddft,martinez1,takatsuka1,takatsuka2} 
which has been regarded as a challenging field for theorists: 
Although most \textit{ab initio} theories are built upon the Born-Oppenheimer approximation   
to separate the nuclear and electronic degrees of freedom, this approximation will break down in the region 
where nonadiabatic transitions occur. In order to describe nonadiabatic processes, it is 
necessary to go beyond the Born-Oppenheimer approximation and take account of nonadiabatic couplings (NAC), 
which is the driving force for nonadiabatic transition to different potential energy surfaces (PES). \cite{mbaer-book} 
Since NAC (preferentially called as first and second 
derivative couplings in quantum chemistry) are defined as matrix elements of the first and second derivatives 
with respect to nuclear coordinates between adiabatic states (many-body wavefunctions), 
nonadiabatic dynamics simulation has long been relying on 
wavefunction-based methods to provide the NAC data.   
For more efficient calculation of NAC, density functional methods, \cite{billeter} 
especially those based on time-dependent density functional theory (TDDFT), 
have been developed in the last decade. 
The study was initiated by Chernyak and Mukamel \cite{chernyak} 
who proposed to perturb the ground state using the  nuclear 
derivative of Hamiltonian and to compute NAC from 
the density response. This scheme was first implemented by Baer \cite{rbaer} 
to study H$_{3}$ using a real-time approach and 
by Hu \textit{et al.} \cite{hu-nac1,hu-nac2} to systematically
study small molecules using the frequency-space formalism of Casida.  
\cite{casida-book,casida96} 
To avoid the pseudopotential problem in the calculation of NAC, all-electron TDDFT schemes 
have been independently developed by Hu \textit{et al.} \cite{hu-ae} and Send \textit{et al.} \cite{furche-nac} 
Alternatively, formulations of NAC from TDDFT has also been achieved by Tavernelli \textit{et al.} \cite{tapavicza, tavernelli,tavernelli-excited} 
using the Casida ansatz, which is promising to correctly give NACs between excited states within the Tamm-Dancoff approximation (TDA). 
A recent study by Hu \textit{et al.} \cite{hu-d-matrix} further clarified relationships between 
different DFT/TDDFT formulations of NAC. \cite{billeter,tapavicza,tavernelli-note,tavernelli-pra}  
These NAC schemes have been applied to nonadiabatic dynamics simulations and have shown that 
TDDFT is promising for balanced cost and performance on the computation of polyatomic systems. 
\cite{tapavicza,tapavicza-jcp,werner,sugino-pccp} 

So far most studies on the computation and application of NAC are focused on the first-order, 
without much discussion on the second-order. Although second-order NAC can be in principle 
expressed by the first-order, the numerical evaluation can not be easily carried out. 
This is because not only the differentiation of first-order NAC  
is needed, but also a complete expansion in eigenstates makes the product of first-order NAC involving 
these states rather complicated. In the wavefunction-based framework, although several methods for evaluating  
second-order NAC have been presented, \cite{redmon,yarkony-2nd,agren} 
there are very few literatures on the direct evaluation of second-order NAC in molecular systems. Correspondingly, 
the practical study by nonadiabatic simulation seldom takes second-order NAC into consideration. 
A simplified procedure is to replace the full quantum description as the quantum-classical simulation, since   
the time evolution of the nuclear degrees of freedom is described by a Poisson bracket that introduces only first order
derivatives. \cite{santer} On the other hand, even the full nonadiabatic operators, 
both first- and second-order NAC, are taken into consideration in the formulation, 
such as \textit{ab initio} multiple spawning, the second-order NAC are just ignored 
in the practice.  \cite{martinez1, martinez2, martinez3} 
It is noted that second-order NAC are often found to be small by experience, \cite{martinez1} however,  
they are not the second-order item in the Taylor expansion but originated from the presence of 
the scalar Laplacian. Therefore, in contrast to the vector form of first-order NAC, the second order 
are scalars. In order to verify the validity of neglecting second-order NAC in nonadiabatic simulations,  
it is crucial to examine the behavior of second-order NAC when the intersection points are approached. 
If the similar diverging behavior as first-order NAC is observed, the neglect of second-order NAC 
needs to be critically reconsidered. 
  
The aim of the present study is to develop an efficient TDDFT method for the calculation of second-order NAC, 
which is desired to have the same-level computational cost as the first-order, and then to examine the behavior of 
second-order NAC near intersection points. For the efficiency, the explicit expansion into first-order NAC should be avoided.  
We will show that this can be achieved by using the Casida ansatz adapted for the first-order NAC, \cite{hu-d-matrix} while 
there is no need to explicitly construct auxiliary excited-state wavefunctions. \cite{tavernelli-note} 
Justification of our procedure can be shown within the TDA. 
To check if the second-order NAC diverge at intersection points, we carry out TDDFT calculations within modified linear 
response theory \cite{hu-pra,hu-jcp} in the immediate vicinity of Jahn-Teller,\cite{bersuker,bersuker-book} Renner-Teller \cite{jungen,mikhailov} 
and elliptic Jahn-Teller intersections, \cite{vibok-NaH2} 
and compare results with model analysis. It is verified that our TDDFT results are in good agreement 
with the predictions from the models. In the vicinity of different types of intersections 
different behaviors of second-order NAC are revealed, either in the Cartesian components ($x$, $y$ and $z$) or as a whole scalar:  
The components  are shown to be negligibly small near Renner-Teller glancing intersections, while they are significantly large 
near the Jahn-Teller conical intersections. Nevertheless, the components of second-order NAC can cancel 
each other to a large extent in Jahn-Teller systems, indicating the background of neglecting 
second-order NAC in practical dynamics simulations.  On the other hand, it is also shown that such a 
cancellation becomes less effective in an elliptic Jahn-Teller system and thus the role of 
second-order NAC needs to be evaluated in the rigorous framework. 

The present paper is organized as follows. In Sec.~II, we present the
formulation of second-order NAC from TDDFT and its extension within modified linear 
response theory. In Sec. III, implementation in the planewave pseudopotential framework 
and computational details are given. In Sec. IV, practical calculations on various
molecular systems possessing Jahn-Teller, Renner-Teller or elliptic Jahn-Teller intersections are
performed, and compared with ideal values predicted by Jahn-Teller / Renner-Teller models. 
In Sec. V, we conclude our work. 

\section{Formulation}
\subsection{Second-order NAC from the adapted Casida ansartz}
In the previous work of Hu \textit{et al.} rigorous TDDFT formulations of first-order NAC have been  
achieved, using the Kohn-Sham matrix elements of either $h$-operator \cite{hu-nac1,hu-nac2,hu-ae} 
or $d$-operator,  \cite{hu-d-matrix} i.e., 
\begin{equation}
\hat{h}_\mu\equiv\frac{\partial \hat{H}}{\partial R_\mu}, \quad \quad  \hat{d}_\mu\equiv\frac{\partial }{\partial R_\mu}, 
\end{equation}  
where $H$ is the many-body Hamiltonian and $R_\mu$ is the nuclear coordinate with $\mu$
representing $x$, $y$, and $z$ components and atom index. 

The $h$-matrix formulation gives first-order NAC as%
\begin{equation}
\left\langle \Psi_{0}\right\vert \frac{\partial}{\partial R_{\mu}}\left\vert
\Psi_{I}\right\rangle =\omega_{I}^{-1}\left\langle \Psi_{0}\right\vert
\frac{\partial\hat{H}}{\partial R_{\mu}}\left\vert \Psi_{I}\right\rangle
=\omega_{I}^{-3/2}\mathbf{h}_{\mu}^{\dagger}\mathbf{S}^{-1/2}\mathbf{F}_{I}.
\label{eq_h_NAC}%
\end{equation}
where $\Psi_{0}$ ($\Psi_{I}$) is the many-body electronic wavefunction of the
ground ($I$-th excited) state, and $\omega_{I}$ is the excitation energy. Matrix
elements of $\mathbf{S}$ and $\mathbf{h}_{\mu}$ are given by%
\begin{equation}
S_{ij\sigma,kl\tau}=\frac{\delta_{\sigma,\tau}\delta_{i,k}\delta_{j,l}%
}{\left(  f_{k\tau}-f_{l\tau}\right)  \left(  \varepsilon_{l\tau}%
-\varepsilon_{k\tau}\right)  }\label{eq_Sij}%
\end{equation}
and%
\begin{equation}
h_{ij\sigma,\mu}=\left\langle \psi_{i\sigma}\right\vert \frac{\partial\hat{H}%
}{\partial R_{\mu}}\left\vert \psi_{j\sigma}\right\rangle ,\label{eq_hij}%
\end{equation}
where $\psi_{i\sigma}$, $\varepsilon_{i\sigma}$, $f_{i\sigma}$ are,
respectively, the orbital, eigenvalue, and occupation number for the $i$-th KS
state with spin $\sigma$. $\mathbf{F}_{I}$ is the eigenvector of the
Casida equation \cite{casida-book}%
\begin{equation}
\mathbf{\Omega F}_{I}=\omega_{I}^{2}\mathbf{F}_{I},\label{eq_casida}%
\end{equation}
where%
\begin{equation} 
\Omega_{ij\sigma,kl\tau}=\delta_{\sigma,\tau}\delta_{i,k}\delta_{j,l}\left(
\varepsilon_{l\tau}-\varepsilon_{k\tau}\right)  ^{2}+2\sqrt{\left(
f_{i\sigma}-f_{j\sigma}\right)  \left(  \varepsilon_{j\sigma}-\varepsilon
_{i\sigma}\right)  }K_{ij\sigma,kl\tau}\sqrt{\left(  f_{k\tau}-f_{l\tau
}\right)  \left(  \varepsilon_{l\tau}-\varepsilon_{k\tau}\right)  }\label{eq_Omega}%
\end{equation}
with $\mathbf{K}$ being the KS matrix of the Hartree and
exchange-correlation (xc) kernel ($\Lambda^{\text{hxc}}$),%
\begin{equation}
K_{ij\sigma,kl\tau}=\int\int d\mathbf{r}d\mathbf{r}^{\prime}%
\psi_{i\sigma}\left(  \mathbf{r}\right)  \psi_{j\sigma}\left(
\mathbf{r}\right)  \Lambda^{\text{hxc}}\left(  \mathbf{r}%
,\mathbf{r}^{\prime}\right)  \psi_{k\tau}\left(  \mathbf{r}^{\prime
}\right)  \psi_{l\tau}\left(  \mathbf{r}^{\prime}\right)  .\label{eq_Kij}%
\end{equation}
The KS orbitals have been assumed to be real for simplicity. 

On the other hand, the $d$-matrix formulation gives first-order NAC as
\begin{equation}
\left\langle \Psi_{0}\right\vert \frac{\partial}{\partial R_{\mu}}\left\vert
\Psi_{I}\right\rangle =\omega_{I}^{1/2}\mathbf{d}_{\mu}^{\dagger
}\mathbf{S}^{1/2}\mathbf{F}_{I},\label{eq_d_NAC}%
\end{equation} 
where 
\begin{equation}
d_{ij\sigma,\mu}=\left\langle \psi_{i\sigma}\right\vert \hat{d}_\mu
\left\vert \psi_{j\sigma}\right\rangle
=
\left\langle \psi_{i\sigma}\right\vert \frac{\partial}
{\partial R_{\mu}}\left\vert \psi_{j\sigma}\right\rangle.
\label{eq_dij}
\end{equation}
The $d$-matrix formulation is derived from the original $h$-matrix formulation,  
using the relationship between the nuclear derivatives of many-body Hamiltonian 
and Kohn-Sham Hamiltonian. It can avoid the problem of the pseudopotential 
approximation in reproducing the inelastic terms corresponding to the 
off-diagonal $h$-matrix elements.  

It is interesting to note that the two TDDFT formulations of first-order 
NAC, Eqs.~(\ref{eq_h_NAC}) and (\ref{eq_d_NAC}), give similar but subtly 
different expressions for the connection between TDDFT quantities and 
many-body theory, i.e., 
\begin{equation}
\mathbf{h}_{\mu}^{\dagger}\mathbf{S}^{-1/2}\mathbf{F}_{I} = \omega_{I}^{1/2}
\left\langle {\Psi}_{0}\right\vert \hat{h}_\mu 
\left\vert {\Psi}_{I}\right\rangle 
\label{eq_h_MBT}
\end{equation} 
and 
\begin{equation}
\mathbf{d}_{\mu}^{\dagger}\mathbf{S}^{1/2}\mathbf{F}_{I} = \omega_{I}^{-1/2}
\left\langle {\Psi}_{0}\right\vert \hat{d}_\mu \left\vert {\Psi}_{I}\right\rangle,  
\label{eq_d_MBT}
\end{equation}
which can be further compared with the one for the dipole operator $\hat{r}_\mu$, 
\begin{equation}
\mathbf{r}_{\mu}^{\dagger}\mathbf{S}^{-1/2}\mathbf{F}_{I} = \omega_{I}^{1/2}
\left\langle {\Psi}_{0}\right\vert \hat{r}_\mu \left\vert {\Psi}_{I}\right\rangle, 
\label{eq_r_MBT} 
\end{equation} 
as derived by Casida for the calculation of oscillator strength.  
\cite{casida-book} 
The expression of the $\hat{d}_\mu$ operator, Eq.~(\ref{eq_d_MBT}), shows a distinct feature as it 
gives different powers in $\mathbf{S}$ and $\omega_I$. 
It is reminded that Eq.~(\ref{eq_r_MBT}) is the basis of the Casida ansatz, 
in which the auxiliary many-body excited-state wavefunction is constructed as  
\begin{equation}
\bar{\Psi}_{I}=\sum_{ij\sigma}^{f_{i\sigma}>f_{j\sigma}}\sqrt{\frac
{\varepsilon_{j\sigma}-\varepsilon_{i\sigma}}{\omega_{I}}}F_{ij\sigma,I}%
\hat{a}_{j\sigma}^{\dagger}\hat{a}_{i\sigma}\bar{\Psi}_{0},\label{eq_ansatz_wav}%
\end{equation} 
so that 
\begin{equation}
\left\langle {\Psi}_{0}\right\vert \hat{O}_\mu \left\vert {\Psi}_{I}\right\rangle 
=\left\langle \bar{\Psi}_{0}\right\vert \hat{O}_\mu \left\vert \bar{\Psi}_{I}\right\rangle.  
\label{eq_ansatz_equiv}
\end{equation} 
Herein $\hat{a}_{j\sigma}^{\dagger}$ and $\hat{a}_{i\sigma}$ are respectively 
creation and annihilation operators, and $\bar{\Psi}_0$ is a Slater determinant 
of occupied KS orbitals.  
Details regarding the Casida ansatz and the mapping between TDDFT quantities and many-body theory  
can be found in Ref.~[\onlinecite{tavernelli-note}].  Nevertheless, 
in order to validate Eq.~(\ref{eq_ansatz_equiv}) 
also for $\hat{O}_\mu$ = $\hat{d}_\mu$, the Casida ansatz need to be adapted 
according to Eq.~(\ref{eq_d_MBT}) in the following way, 
\begin{equation}
\tilde{\Psi}_{I}=\sum_{ij\sigma}^{f_{i\sigma}>f_{j\sigma}}\sqrt{\frac
{\omega_{I}}{\varepsilon_{j\sigma}-\varepsilon_{i\sigma}}}F_{ij\sigma,I}%
\hat{a}_{j\sigma}^{\dagger}\hat{a}_{i\sigma}\tilde{\Psi}_{0},  
\label{eq_ansatz_nac}
\end{equation}
where $\tilde{\Psi}_0 = \bar{\Psi}_0$. 

With the adapted Casida ansatz in hand, we can now readily derive the second-order NAC, 
assuming the similarity between first- and second-derivative operators. 
Defining   
\begin{equation}
\hat{b}_\mu\equiv\frac{\partial^2 }{\partial R_\mu^2}, 
\end{equation}  
we can get  
\begin{equation}
\left\langle {\Psi}_{0}\right\vert \hat{b}_\mu \left\vert {\Psi}_{I}\right\rangle 
=\left\langle \tilde{\Psi}_{0}\right\vert \hat{b}_\mu \left\vert \tilde{\Psi}_{I}\right\rangle.  
\label{eq_ansatz_equiv_b}
\end{equation} 
from the adapted Casida ansatz. 
Since Eq.~(\ref{eq_ansatz_nac}) is equivalent to 
\begin{equation}
\tilde{\Psi}_{I}=\sum_{ij\sigma}^{f_{i\sigma}>f_{j\sigma}}
\omega_I^{1/2}\left(\mathbf{S}^{1/2}\mathbf{F}_I\right)_{ij\sigma} 
\hat{a}_{j\sigma}^{\dagger}\hat{a}_{i\sigma}\tilde{\Psi}_{0}, 
\label{eq_ansatz_wav}%
\end{equation}
further using the connection from the Casida ansatz to  
the mapping between TDDFT and many-body theory \cite{tavernelli-note}, we can get 
\begin{equation}
\mathbf{b}_{\mu}^{\dagger}\mathbf{S}^{1/2}\mathbf{F}_{I} = \omega_{I}^{-1/2}
\left\langle {\Psi}_{0}\right\vert \hat{b}_\mu \left\vert {\Psi}_{I}\right\rangle,  
\label{eq_b_MBT}
\end{equation} 
i.e., 
\begin{equation}
\left\langle {\Psi}_{0}\right\vert \frac{\partial ^2}{\partial R_\mu ^2} \left\vert {\Psi}_{I}\right\rangle
=\omega_{I}^{1/2}\mathbf{b}_{\mu}^{\dagger}\mathbf{S}^{1/2}\mathbf{F}_{I}. 
\label{eq_b_2nd_NAC}
\end{equation}  
This expression shows that we can calculate second-order NAC without explicitly 
constructing (auxiliary) excited wavefunctions. Moreover, it is appealing that the computational cost of 
second-order NAC by this expression is at the same-level as that of the first-order. 
On the other hand, it is noted that although the derivation 
of Eq.~(\ref{eq_d_MBT}) is rigorous,  
derivation of Eq.~(\ref{eq_b_MBT}) is not yet.  
The validity of the adapted Casida ansatz for the second-order 
NAC needs to be further justified. Next we show that this can be achieved  
within the TDA, where the adapted Casida ansatz 
becomes equivalent to the original one.  

\subsection{Second-order NAC within the TDA}
The justification of the second-order NAC formulation can be attempted by using the expansion of first-order NAC to 
show the validity of Eq.~(\ref{eq_ansatz_equiv_b}). 
It has been shown \cite{hu-d-matrix,tavernelli-excited} that for those between ground state and singly excited states, it generally holds that 
\begin{equation}
\left\langle {\Psi}_{0}\right\vert \hat{d}_\mu \left\vert {\Psi}_{I}\right\rangle 
=\left\langle \tilde{\Psi}_{0}\right\vert \hat{d}_\mu \left\vert \tilde{\Psi}_{I}\right\rangle,
\label{eq_ansatz_equiv_d}
\end{equation}
and for those between singly excited states, the validity of the expression 
\begin{equation}
\left\langle {\Psi}_{I}\right\vert \hat{h}_\mu \left\vert {\Psi}_{J}\right\rangle 
=\left\langle \bar{\Psi}_{I}\right\vert \hat{h}_\mu \left\vert \bar{\Psi}_{J}\right\rangle
\label{eq_ansatz_equiv_h}
\end{equation}
can be justified using the TDA, where 
$\omega_{I}^{1/2}\mathbf{S}^{1/2}=1$, and the two forms of auxiliary wavefunctions become 
the same, i.e., $\bar{\Psi}_I=\tilde{\Psi}_I$. 
The second-order NAC can be expanded by the first-order as 
\begin{multline}
\left\langle {\Psi}_{0}\right\vert \hat{b}_\mu \left\vert {\Psi}_{I}\right\rangle
=-\langle\frac{\partial}{\partial R_\mu}\Psi _0 | \frac {\partial}{\partial R_\mu}\Psi _I \rangle + \frac {\partial}{\partial R_\mu}
\left\langle {\Psi}_{0}\right\vert \frac{\partial }{\partial R_\mu} \left\vert {\Psi}_{I}\right\rangle \\
=-\sum_m \langle\frac{\partial}{\partial R_\mu}\Psi _0 | \Psi _m\rangle \langle\Psi _m | \frac{\partial}{\partial R_\mu}\Psi _I\rangle
+\frac{\partial}{\partial R_\mu}
\left\langle {\Psi}_{0}\right\vert \frac{\partial }{\partial R_\mu} \left\vert {\Psi}_{I}\right\rangle \\
=\sum_m \left\langle {\Psi}_{0}\right\vert \hat{d}_\mu \left\vert {\Psi}_{m}\right\rangle
\frac{\left\langle {\Psi}_{m}\right\vert \hat{h}_\mu\left\vert {\Psi}_{I}\right\rangle}{E_I-E_m}
+\frac{\partial}{\partial R_\mu}
\left\langle {\Psi}_{0}\right\vert \hat{d}_\mu\left\vert {\Psi}_{I}\right\rangle,  
\label{eq_true_expansion}
\end{multline}
which rigorously holds since $\langle \Psi _m | \Psi _ n \rangle = \delta _{mn}$. 
Similarly, if we can show this orthonormalized condition for the auxiliary wavefunction, i.e., 
$\langle \tilde{\Psi} _m | \tilde{\Psi}_ n \rangle = \delta _{mn}$, we can get 
\begin{multline}
\left\langle \tilde{\Psi}_{0}\right\vert \hat{b}_\mu \left\vert \tilde{\Psi}_{I}\right\rangle
=-\langle\frac{\partial}{\partial R_\mu}\tilde{\Psi} _0 | \frac {\partial}{\partial R_\mu}\tilde{\Psi} _I \rangle + \frac {\partial}{\partial R_\mu}
\left\langle \tilde{\Psi}_{0}\right\vert \frac{\partial }{\partial R_\mu} \left\vert \tilde{\Psi}_{I}\right\rangle \\
=-\sum_m \langle\frac{\partial}{\partial R_\mu}\tilde{\Psi} _0 | \tilde{\Psi} _m\rangle \langle\tilde{\Psi} _m | \frac{\partial}{\partial R_\mu}\tilde{\Psi} _I\rangle
+\frac{\partial}{\partial R_\mu}
\left\langle \tilde{\Psi}_{0}\right\vert \frac{\partial }{\partial R_\mu} \left\vert \tilde{\Psi}_{I}\right\rangle \\
=\sum_m \left\langle \tilde{\Psi}_{0}\right\vert \hat{d}_\mu \left\vert \tilde{\Psi}_{m}\right\rangle
\frac{\left\langle \tilde{\Psi}_{m}\right\vert \hat{h}_\mu\left\vert \tilde{\Psi}_{I}\right\rangle}{E_I-E_m}
+\frac{\partial}{\partial R_\mu}
\left\langle \tilde{\Psi}_{0}\right\vert \hat{d}_\mu\left\vert \tilde{\Psi}_{I}\right\rangle. 
\label{eq_aug_expansion}
\end{multline}
From Eq.~(\ref{eq_ansatz_equiv_d}) and Eq.~(\ref{eq_ansatz_equiv_h}), the identity of Eq.~(\ref{eq_true_expansion}) and Eq.~(\ref{eq_aug_expansion}) can be justified 
provided that  $\tilde{\Psi}_I=\bar{\Psi}_I$, since we have to reconstruct the auxiliary wavefunction from $\tilde{\Psi}_m$ to $\bar{\Psi}_m$ when the operator is changed 
from $\hat{d}_\mu$ to $\hat{h}_\mu$. 
This is satisfied when the TDA is valid. In the meanwhile, 
the orthonormalized condition that 
\begin{multline}
\delta _{IJ}=\langle \tilde{\Psi}_I | \tilde{\Psi}_j \rangle 
= \sqrt{\omega_I \omega_J} \sum_{ij\sigma} \sum_{kl\tau} 
\left( \mathbf{S}^{1/2}\mathbf{F}_I \right)^{\dagger}_{ij\sigma} \left( \mathbf{S}^{1/2}\mathbf{F}_I \right)_{kl\tau} 
\left\langle \tilde{\Psi}_{0}\right\vert \hat{a}^{\dagger}_{i\sigma}\hat{a}_{j\sigma}\hat{a}^{\dagger}_{l\tau}\hat{a}_{k\tau} \left\vert \tilde{\Psi}_{0}\right\rangle
\\
= \sqrt{\omega_I \omega_J} \sum_{ij\sigma} \sum_{kl\tau} 
\left( \mathbf{S}^{1/2}\mathbf{F}_I  \right)^{\dagger}_{ij\sigma} \left( \mathbf{S}^{1/2}\mathbf{F}_I \right)_{kl\tau} 
\delta_{ik}\delta_{jl}\delta_{\sigma\tau}
= \sqrt{\omega_I \omega_J} \mathbf{F}^{\dagger}_I \mathbf{S}\mathbf{F}_J 
\end{multline}
also holds within the TDA since $\mathbf{F}^{\dagger}_I \mathbf{F}_J = \delta_{IJ}$. 
Therefore, Eq.~(\ref{eq_true_expansion}) and Eq.~(\ref{eq_aug_expansion}) 
become identical, i.e., the validity of Eq.~(\ref{eq_ansatz_equiv_b}) is justified within the TDA. 

 Further remark is on the complete expansion in Eq.~(\ref{eq_aug_expansion}). As long as we only consider a singly excited state, 
 this does not pose a problem since $\tilde{\Psi}_0$ is a single Slater determinant and only the contributions from other  
singly excited states enter the expansion. 

\subsection{Extension within TDDFT modified linear response theory: Justification of the Slater transition state method}

In the calculation of first-order NAC, a particular example is the case of the Slater transition state method 
for doublet systems.  
Billeter and Curioni \cite{billeter} has used the following expression,  
\begin{equation}
\langle\Psi_0|\frac{\partial}{\partial R_\mu}|\Psi_I\rangle=\langle\psi_{i\sigma}^m|\frac{\partial}{\partial R_\mu}|\psi_{j\sigma}^m\rangle, 
\end{equation}  
where the ($i$,$j$) pair is the particle-hole orbitals responsible for the $I$-th transition, 
and $m$ denotes the mid-excited state (Slater transition state) in which the particle-hole orbitals 
are each filled with a half electron. They have found that this expression can give accurate results 
of first-order NAC between doublet states of molecules at equilibrium geometries, and their approach 
is further validated by our TDDFT modified linear response theory \cite{hu-pra,hu-jcp} and 
also by our calculations near intersection points. \cite{hu-d-matrix}
Next we will show that the extension of TDDFT formulation of second-order NAC  
within modified linear response theory,  
is also equivalent to the Slater transition state method for doublet systems.

Within modified linear response, the excitation energy is calculated from the response of 
the mid-excited state, while other terms in the NAC formula are calculated from that of the pure-state 
configuration. \cite{hu-jcp} Corresponding to the mid-excited state of a doublet system, the adapted Casida equation, 
\begin{equation}
\mathbf{\Omega}^m\mathbf{F}_I^m=\omega_I^m\mathbf{F}_I^m
\end{equation}
with the matrix element   
\begin{equation}
\Omega_{ij\sigma,kl\tau}^m=\delta_{i,k}\delta_{j,l}\delta_{\sigma,\tau}(\epsilon_{j\sigma}^m-\epsilon_{i\sigma}^m)^2
+2(f_{i\sigma}^m-f_{j\sigma}^m)(\epsilon_{j\sigma}^m-\epsilon_{i\sigma}^m) K_{ij\sigma,kl\tau}^m,
\end{equation} 
 gives 
\begin{equation}
\omega_I^m=\epsilon_{j\sigma}^m-\epsilon_{i\sigma}^m, 
\end{equation}
since $f_{i\sigma}^m$=$f_{j\sigma}^m$ = 0.5 in the mid-excited state of a doublet system, which renders the 
corresponding off-diagonal elements of $\mathbf{\Omega}$ to be zero. 
On the other hand, the pure state configuration in the mid-excited state, which 
uses the occupation number of the ground state while 
keeping other quantities of the mid-excited state, gives 
\begin{equation}
\mathbf{b}_{\mu,p}^\dagger\mathbf{S}_p^{1/2}\mathbf{F}_I^p=b_{ij\sigma}^m(\epsilon_{j\sigma}^m-\epsilon_{i\sigma}^m)^{-1/2}, 
\end{equation}
due to the fact that $F_{ij\sigma,I}^p$ is practically equivalent to 1 and other components of $\mathbf{F}_I$ are zero.  
Therefore, 
\begin{equation}
\langle\Psi_0|\hat{b}_\mu|\Psi_I\rangle
=(\omega_I^m)^{1/2}\mathbf{b}_{\mu,p}^\dagger\mathbf{S}_p^{1/2}\mathbf{F}_I^p
=b_{ij\sigma}^m=\langle\psi_{i\sigma}^m|\frac{\partial ^2}{\partial R_\mu ^2}|\psi_{j\sigma}^m\rangle,
\end{equation} 
which is just the second-derivative coupling matrix element between the particle-hole orbitals. 
As a result, the TDDFT formulation of second-order NAC in doublet systems is just reduced to 
the Slater transition state method. 

\section{Implementation and computational details}
The implementation of the present TDDFT method for second-order NAC 
is based on the \textsc{ABINIT} code, \cite{abinit} 
which is a planewave pseudopotential approach.  
All calculations are performed within adiabatic LSDA using the Teter Pade parametrization. \cite{teter-pade} 
The Troullier-Martins pseudopotentials \cite{tm} with nonlinear core correction, \cite{louie} generated 
by Khein and Allan, 
are used for various atomic species.
Only the $\Gamma$ point ($k$=0) is taken into consideration 
in the $\mathbf{k}$ point sampling, which corresponds to the use of real wavefunctions. 
Convergence parameters, such as the supercell size, number of unoccupied orbitals, and kinetic 
energy cutoff, are examined to ensure reasonably accurate results. 
On the basis of the previous implementation of modified linear response theory  
in \textsc{ABINIT},\cite{hu-jcp} its extension for calculating second-order NAC  
requires almost no additional labor, 
since it is only necessary to construct the pure-state configuration from the mid-excited 
state, and to apply the same calculation procedures as ordinary linear response theory.  
To check the performance of our method, it is desired to compare TDDFT results for general atomic geometries 
with those from wavefunction-based methods, however, there are too few literatures on this aspect 
and the direct comparison is difficult. Therefore, we concentrate on evaluating second-order NAC 
in the immediate vicinity of Jahn-Teller, Renner-Teller and elliptic Jahn-Teller intersections, 
where we can directly compare our results with predictions from corresponding models. 

\subsection*{Finite difference method of calculating $b$-matrix elements}
The calculation of $b$-matrix elements is implemented in a straightforward finite-difference scheme, 
with the consideration of aligning the phases of KS orbitals, \cite{billeter} 
as shown by 
\begin{equation}
\langle\psi_{i\sigma}|\hat{b}_\mu|\psi_{j\sigma}\rangle
=\frac{\langle\psi_{i\sigma}(\mathbf{R})|\psi_{j\sigma}(\mathbf{R}+\Delta R\cdot\mathbf{e}_\mu)
\mathrm{sgn}(\xi_{+})
-2\psi_{j\sigma}(\mathbf{R})
+\psi_{j\sigma}(\mathbf{R}-\Delta R\cdot\mathbf{e}_\mu)
\mathrm{sgn}(\xi_{-})\rangle}{\Delta R}, 
\end{equation}
where $\mathbf{e}_\mu$ is the unit vector along the $\mu$ axis, 
$\mathrm{sgn}(\xi)$ is the sign function, i.e., 
\begin{equation}
\mathrm{sgn}(\xi)=\left\{
\begin{array}{rl}
-1 & \textrm{if $\xi<0$} \\
 1 & \textrm{if $\xi>0$}
\end{array}
 \right.
\end{equation}
and 
\begin{equation}
\xi_+=\langle\psi_{j\sigma}(\mathbf{R})|\psi_{j\sigma}(\mathbf{R}+\Delta R\cdot\mathbf{e}_\mu)\rangle, 
\end{equation}
\begin{equation}
\xi_-=\langle\psi_{j\sigma}(\mathbf{R})|\psi_{j\sigma}(\mathbf{R}-\Delta R\cdot\mathbf{e}_\mu)\rangle.
\end{equation}
The accuracy of the above numerical differentiation scheme is checked by using different 
$\Delta R$. In the practice, we choose $\Delta R$=0.002$\sim$0.004 bohr. 

\section{Results and Discussions\label{result}} 
In this section, we present calculation results on various molecular systems possessing Jahn-Teller, Renner-Teller,  
or elliptic Jahn-Teller intersections, where the ground state and the first excited 
state of these molecular systems are degenerate.

\subsection{Jahn-Teller systems}

\begin{figure}
\includegraphics[width=5.0cm]{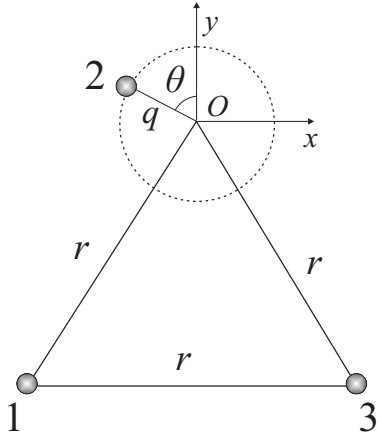}%
\caption{The geometry of the X$_3$ system as one X atom (numbered as 2) 
is moved on the contour around the intersection point (located at $O$). The nuclear 
configuration at the intersection point is an equilateral triangle with $C_{3v}$ 
symmetry, corresponding to the degeneracy of the ground state and the first 
excited state. 
\label{fig_X3-contour}
}
\end{figure}

\begin{figure}
\includegraphics[width=8.5cm]{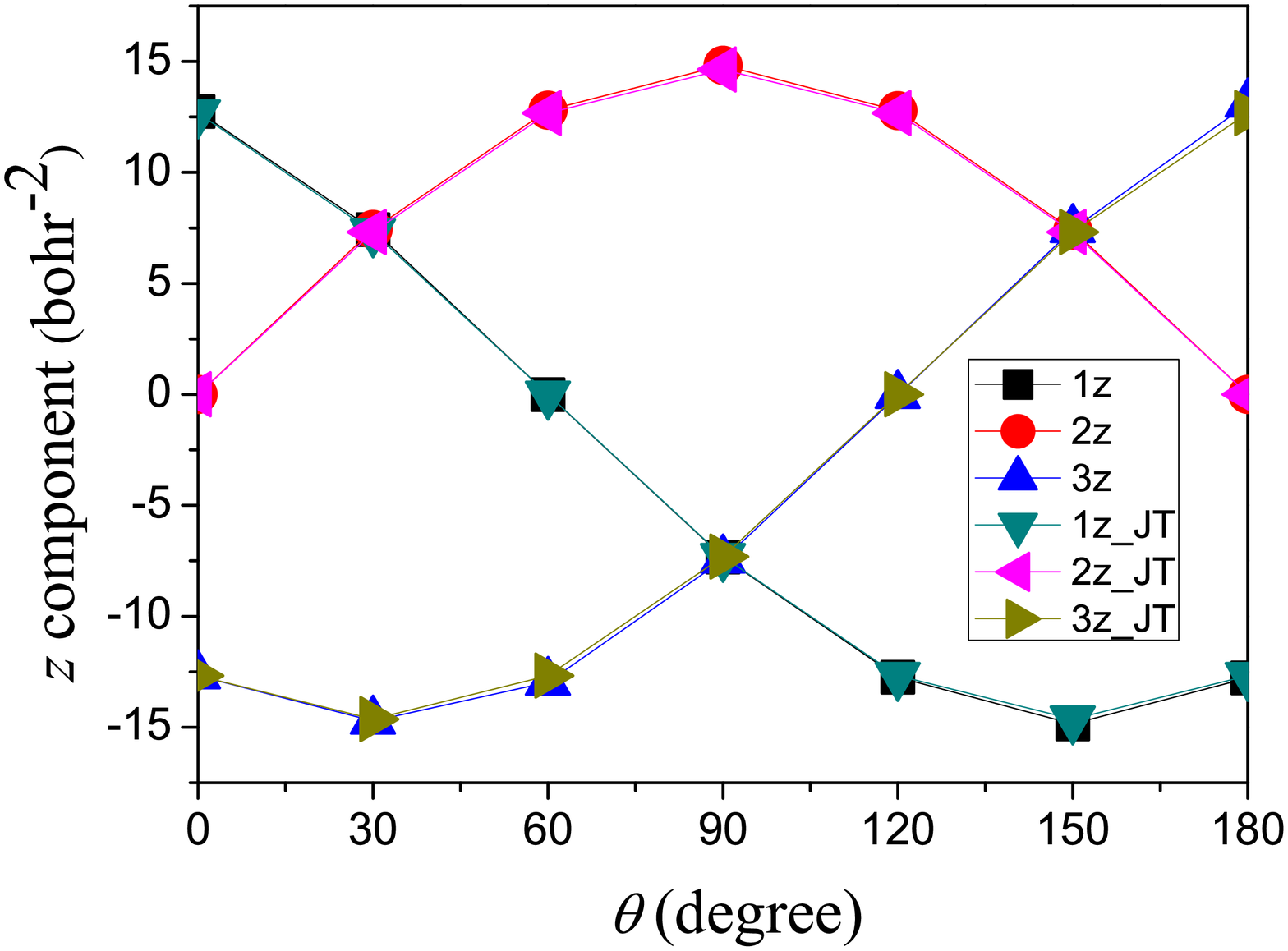}%
\caption{The $z$ components of second-order NAC on the three atoms of H$_3$. 
The labels 1z, 2z and 3z denote the $z$ components on atom 1, 2 and 3, while 
1z{\_}JT, 2z{\_}JT and 3z{\_}JT denote those from the Jahn-Teller model, respectively.    
\label{fig_2ndNAC_H3_z}
}
\end{figure}

In Table~\ref{tab_JT_trimer} we list the $x$, $y$, and $z$ components of second-order NAC in two typical Jahn-Teller systems: 
The prototype H$_3$ molecule \cite{abrol,mbaer-cpl,mbaer-H3} and an alkali-metal trimer Li$_3$. \cite{martins-Li3Na3K3,gerber} The three 
atoms are located in the geometry of Fig.~\ref{fig_X3-contour} in which one atom is moved on the contour around the intersection point. 
The contour radius $q$ is chosen as 0.02 bohr, which is sufficiently small so as to 
be comparable to the condition of the Jahn-Teller model. 
It is clearly seen that at such a small $q$  the $x$ and $y$ components of second-order NAC in H$_3$ and Li$_3$ are 
significantly large: The nonzero values are in the order of 1000 bohr$^{-2}$, which are much larger than those of 
first-order NAC (which are in the order of $1/q$). In the meanwhile, both the magnitude and relative signs of TDDFT results are in 
good agreement with the Jahn-Teller model. (The ideal values of second-order NAC from the Jahn-Teller model 
can be derived from the derivatives of the first-order NAC, as shown by Appendix~\ref{app_JT}.) 
On the other hand, it is noted that the $z$ components of second-order NAC are quite small but nonzero, either in H$_3$ or Li$_3$. 
This is different from the zero values of $z$ components of first-order NAC in X$_3$ systems near intersection points.  
The Jahn-Teller model predicts that the $x$ and $y$ components of second-order NAC only depend on contour radius $q$ 
and angle $\theta$, while there is an additional dependence of $z$ component on the internuclear distance $r$. 
In our calculations we set $r_\mathrm{H-H}$ = 1.9729 bohr 
and $r_\mathrm{Li-Li}$ = 5.0 bohr respectively, therefore,  TDDFT calculations are expected 
to give different $z$ components for H$_3$ and Li$_3$, and the results seem to give such a difference: Both the magnitude and sign 
agree with the ideal values corresponding to the above internuclear distances. However, since the $z$ components 
are quite small we need to make sure whether they are intrinsically nonzero. For this purpose we have made a detailed 
examination of the $z$ components of second-order NAC on the three atoms of H$_3$ as a function of the contour angle $\theta$, 
as shown by Fig.~\ref{fig_2ndNAC_H3_z}. As $\theta$ is varied from 0 to 180$^\circ$, the $z$ components on all atoms, although small, 
show clear dependences on $\theta$ and agree well with the Jahn-Teller model. This means that the small 
$z$ components of second-order NAC are intrinsically nonzero and have been accurately reproduced by TDDFT. 

\renewcommand\arraystretch{0.98}
\begin{table}
\caption{\label{tab_JT_trimer}The calculated $x$, $y$ and $z$ components of second-order NAC (in bohr$^{-2}$) 
on three atoms of  
H$_3$ and Li$_3$, which are at the geometry of Fig.~\ref{fig_X3-contour}.   
The contour radius $q$ is 0.02 bohr and angle $\theta$ is 0. The ideal values 
from the Jahn-Teller model, as derived in Appendix~\ref{app_JT} and summarized in Table~\ref{tab_JT_2nd_NAC}, 
are also listed for comparison.  It is noted that the $z$ components of 
second-order NAC in the Jahn-Teller model are dependent on atomic distances and have been derived within 
two sets of parameters:  the values inside the parenthesis are derived by $r_\text{Li-Li}$ = 5.0 bohr, 
while the others outside are derived by $r_\text{H-H}$ = 1.9729 bohr. 
}
\begin{ruledtabular}
\begin{tabular}{llrrr} 
 &  & $x\quad$ & $y\quad$ & $z\quad$\\
\hline
H$_3$  & atom 1 & 1085.88  & -1074.36 &  12.75 \\
       & atom 2 &  0.30    &  0.00    &  0.00 \\
       & atom 3 & -1085.30 &  1073.36 & -12.68 \\
\\
Li$_3$ & atom 1 & 1102.08 & -1090.50 &  4.37 \\
       & atom 2 &  1.44 &   0.00     & 0.00 \\
       & atom 3 & -1099.12 &  1086.05 & -4.71 \\ 
\\
Model & atom 1 & 1082.53 & -1082.53 & 12.67 (5.0) \\
      & atom 2 &   0.00 &   0.00 &    0.0 (0.0)  \\
      & atom 3 & -1082.53 & 1082.53 & -12.67 (-5.0)  \\    
\end{tabular}
\end{ruledtabular} 
\end{table}

Another point noteworthy in Table~\ref{tab_JT_trimer} is the sum of $x$, $y$ and $z$ components. 
In contrary to the vector form of first-order NAC, second-order NAC are scalars due to the presence of 
the scalar Laplacian, therefore, only the sum of $x$, $y$ and $z$ components are meaningful in the 
nonadiabatic dynamics simulation. Table~\ref{tab_JT_trimer} shows the sum of components in both H$_3$ 
and Li$_3$ are small,  as predicted by the Jahn-Teller model. This can provide the background for the 
the neglect of second-order NAC in practical simulations. \cite{martinez1,martinez2,martinez3}

\subsection{Renner-Teller systems}

\begin{figure}
\includegraphics[width=8.0cm]{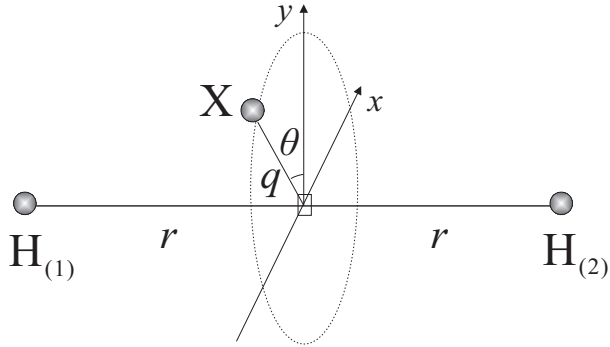}%
\caption{Geometry of the XH$_2$ or XH$_2^+$ system when 
the X atom is moved on the contour around the Renner-Teller intersection point 
(indicated by the open square) on the collinear axis. 
The contour, with radius $q$  and angle $\theta$, 
is fixed in the $xy$ plane, which is perpendicular to the HH axis. 
The two hydrogen atoms are set to be symmetric to the plane.  
\label{fig_RT_XH2} 
}
\end{figure}

In Table~\ref{tab_RT_trimer} we list the $x$, $y$, and $z$ components of second-order NAC in several typical 
Renner-Teller systems: The BH$_2$, CH$_2^+$, NH$_2$ and  H$_2$O$^+$ molecules, \cite{liu,halasz-NH2} which 
are in the geometry of Fig.~\ref{fig_RT_XH2} with the contour angle $\theta$ = 0. 
The contour radius $q$ is chosen as 0.1 bohr, which is known to be sufficiently small and can   
be comparable to the condition of the Renner-Teller model. 
The internuclear distances are set as $r_{\mathrm{H-B}}$ = 2.0 bohr,
$r_{\mathrm{H-C}}$ = 2.0 bohr,  $r_{\mathrm{H-N}}$ = 1.95 bohr, and $r_{\mathrm{H-O}}$ = 1.85 bohr. 
It is interesting to see that all components of second-order NAC on three atoms of all molecules  
are negligibly small (reminding that the first-order NAC in Renner-Teller systems are in the order of 
$1/q$) and agree with the Renner-Teller model.  (The ideal values of second-order NAC from the Renner-Teller model 
can be derived from the derivatives of the first-order NAC, as shown by Appendix~\ref{app_RT}.) 
As a matter of fact, NAC of Renner-Teller system do not dependent on the contour angle $\theta$, and 
thus the negligibly small values of second-order NAC in demonstrated systems are not accidental results 
for a specified geometry, but indicate that they are intrinsically zero. 
In connection with the nonadiabatic dynamics simulation, it is thus verified that the sum of $x$, $y$ and $z$ components can be 
regarded as zero in Renner-Teller systems. This also provides a background for the 
the neglect of second-order NAC in practical dynamics simulations. \cite{martinez1,martinez2,martinez3} 

\begin{table}
\caption{\label{tab_RT_trimer}The calculated $x$, $y$ and $z$ components of second-order NAC (in bohr$^{-2}$) 
on three atoms of  
BH$_2$, CH$_2^+$, NH$_2$ and H$_2$O$^+$, which are at the geometry of Fig.~\ref{fig_RT_XH2}.   
The contour radius $q$ is 0.1 bohr and the angle $\theta$ is 0. The ideal values 
from the Renner-Teller model, as derived in Appendix~\ref{app_RT}, are also listed for comparison. } 
\begin{ruledtabular}
\begin{tabular}{llrrr} 
 &  & $x\quad$ & $y\quad$ & $z\quad$\\
\hline
BH$_2$ & atom H$_{(1)}$ & 0.015 &  0.0 & 0.0 \\
       & atom B  & 0.016 &  0.0 & 0.0 \\
       & atom H$_{(2)}$ & 0.015 &  0.0 & 0.0 \\       
\\
CH$_2^+$ & atom H$_{(1)}$ & -0.012 &  0.0 & 0.0 \\
         & atom C         &  0.013 &  0.0 & 0.0 \\
         & atom H$_{(2)}$ & -0.012 &  0.0 & 0.0 \\       
\\
NH$_2$ & atom H$_{(1)}$ & 0.018 & 0.0 & 0.0 \\
       & atom N  & 0.018 & 0.0 & 0.0 \\
       & atom H$_{(2)}$ & 0.018 & 0.0 & 0.0 \\ 
\\
H$_2$O$^+$ & atom H$_{(1)}$ & -0.0035 &  0.0 & 0.0 \\
           & atom O         &  0.021 &  0.0 & 0.0 \\
           & atom H$_{(2)}$ &  0.0014 &  0.0 & 0.0 \\       
\\
Model & atom 1 & 0.0 & 0.0 & 0.0 \\
      & atom 2 & 0.0 & 0.0 & 0.0 \\
      & atom 3 & 0.0 & 0.0 & 0.0 \\    
\end{tabular}
\end{ruledtabular} 
\end{table}

\subsection{Elliptic Jahn-Teller system}

\begin{figure}
\includegraphics[width=5.0cm]{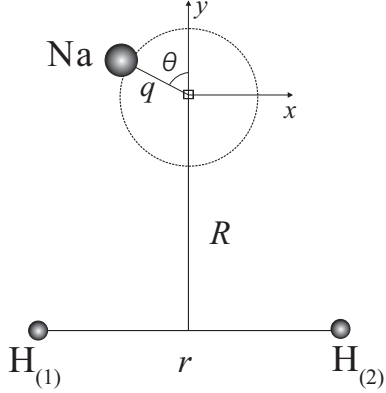}%
\caption{The geometry of the NaH$_2$ system as the Na atom is moved 
on the contour around the conical intersection point (indicated by the open 
square). The nuclear configuration at the conical intersection point is an 
isoseles triangle with C$_{2v}$ symmetry. 
\label{fig_NaH2_contour}
}
\end{figure}

\begin{table}
\caption{\label{tab_NaH2_2nd_NAC}Calculated values of $x$, $y$ and $z$ components of second-order NAC (in bohr$^{-2}$) on the  
three atoms of NaH$_2$, which are in the geometry shown by Fig.~\ref{fig_NaH2_contour} with the 
contour radius $q$=0.1 bohr and angle $\theta$ = 60$^\circ$.  
}
\begin{ruledtabular}
\begin{tabular}{lccc} 
 & $x$ & $y$ & $z$ \\
\hline
atom Na        &  20.54   &  -110.13  &   1.23  \\
atom H$_{(1)}$ & -542.53  &  -10.94   &  -2.89  \\
atom H$_{(2)}$ & -618.23  &   51.59   &  -5.42  \\   
\end{tabular}
\end{ruledtabular} 
\end{table} 

In Table~\ref{tab_NaH2_2nd_NAC} we list TDDFT calculation results of the $x$, $y$, and $z$ components of second-order NAC on the three 
atoms of NaH$_2$, which is known as an elliptic Jahn-Teller system.\cite{yarkony-NaH2,vibok-NaH2} 
The three atoms are located in the geometry of Fig.~\ref{fig_NaH2_contour} with the 
contour angle $\theta$ = 60$^\circ$.  Other parameters regarding the geometry are $r$ = 2.18 bohr  
and $R$ = 3.6127 bohr, according to the intersection point determined in our previous work.\cite{hu-nac2}
For an elliptic Jahn-Teller systems, the angular NAC $A_\theta$ is not just in a quantized value of 0.5, 
but shows a strong dependence on the contour angle $\theta$. Using the similar procedures in Appendix~\ref{app_JT} 
but setting the angular NAC, $A_\theta$, as a variable rather than a constant of 0.5, we can easily get the conclusion 
that second-order NAC would depend on $\frac{\partial A_\theta}{\partial \theta}$, i.e., the slope of 
angular NAC with respect to $\theta$. Therefore,  
we set $\theta$ as 60$^\circ$ at which $\frac{\partial A_\theta}{\partial \theta}$  
is relatively large, as revealed by our previous work.\cite{hu-nac2}
It is clearly seen that under such a condition the magnitudes of $x$ and $y$ components  
become unbalanced in comparison with the Jahn-Teller systems. Meanwhile, the $z$ components are still relatively small. 
Therefore, the sum of $x$, $y$ and $z$ components of second-order NAC can not be negligibly small. 
Regarding the role of NAC in the nonadiabatic dynamics simulation, it is thus suggested to 
include second-order NAC in the rigorous simulation of general molecular systems, which might possess 
accidental conical intersections without any symmetry requirements as in Jahn-Teller systems. 

\section{Conclusion\label{conclude}} 
We have proposed an efficient TDDFT method for calculating second-order NAC between ground state and singly 
excited states, which is based on the Casida ansatz 
adapted for first-order NAC, while the calculation procedure can be done without the need of explicitly constructing auxiliary excited-state 
wavefunctions. Our formulation can be justified when the TDA is valid. 
Within the modified linear response theory, the TDDFT formulation is reduced to the Slater transition method 
in doublet systems. Test calculations are carried out in the immediate vicinity of various types of intersection points. 
The results are in good agreement with the ideal values derived from Jahn-Teller or Renner-Teller models. 
Contrary to the diverging behavior of first-order NAC near 
intersections, the Cartesian components of second-order NAC are shown to be negligibly small near Renner-Teller 
intersections, while they are significantly large near Jahn-Teller intersections. Nevertheless, the Cartesian components 
of second-order NAC can cancel each other to a large extent in Jahn-Teller systems, showing the 
background of neglecting second-order NAC in nonadiabatic dynamics simulations. On the other hand, it is 
revealed that such a cancellation becomes less effective in an elliptic Jahn-Teller system and 
thus the role of second-order NAC needs to be evaluated in the rigorous framework.  
Finally, it is noted that the performance of TDDFT on the computation of second-order NAC needs to be further validated, 
particularly for a general atomic geometry, which requires reference data and remains future work.  

\begin{acknowledgments} 
The authors thank Dr. Yoshitaka Tateyama, Mr. Jun Haruyama, and Mr. Yohei Iwami
for fruitful discussions. 
This work was supported in part by the Project of Materials Design through Computics: Complex Correlation and Non-Equilibrium Dynamics, 
a Grant in Aid for Scientific Research on Innovative Areas, 
and the Next Generation Super Computing Project, Nanoscience Program, MEXT, Japan. 
C. H. thanks the support by State Key Laboratory of New Ceramic and Fine Processing, Tsinghua University.
K. W. acknowledges partial financial support from MEXT through a Grant-in-Aid (No. 19540411 and No. 22104007).   
Testing of our program has been performed on the supercomputers of Institute for Solid State Physics, 
University of Tokyo.  
\end{acknowledgments}

\appendix

\section{\label{app_JT}Second-order NAC from the Jahn-Teller model} 
\begin{figure}
\includegraphics[width=5.0cm]{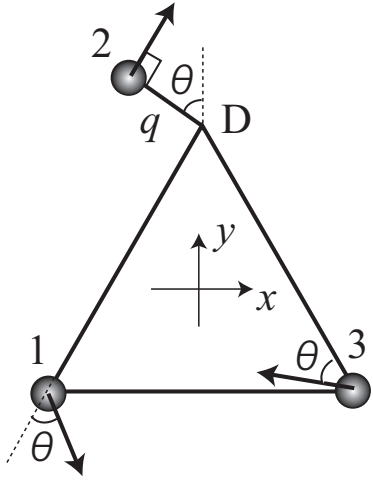}%
\caption{Configuration of a Jahn-Teller trimer in the $xy$ plane, in which atom 2 is regarded on a 
contour with radius $q$ and angle $\theta$ around the intersection 
point (vertex D of the equilateral triangle). Arrows represent NAC vectors 
on the three atoms.  
\label{fig_JT_1st_NAC}
}
\end{figure}

The Jahn-Teller model describes a class of systems in which a set of nuclear coordinates are 
coupled to a two-level system consisting of the ground state and the first excited state of 
appropriate symmetry \cite{lorquet}. 
Figure~\ref{fig_JT_1st_NAC} shows an arbitrary configuration of a Jahn-Teller trimer. 
When the contour radius $q$ is sufficiently small,  the angular NAC has a quantized value of $\frac{1}{2}$ according to the 
Jahn-Teller model. \cite{lorquet} All components of first-order NAC on the three atoms can thus be uniquely determined, as shown by 
Table~\ref{tab_JT_1st_NAC}.  

\begin{table}
\caption{\label{tab_JT_1st_NAC}The $x$, $y$ and $z$ components of first-order NAC on 
the three atoms of a Jahn-Teller trimer, which are in the geometry shown by Fig.~\ref{fig_JT_1st_NAC}.  
$q$ is the contour radius and $\theta$ is the contour angle. 
}
\begin{ruledtabular}
\begin{tabular}{lccc} 
 & $x$ & $y$ & $z$ \\
\hline
atom 1 & $\frac{0.5}{q}\cos(120^\circ-\theta)$  & $-\frac{0.5}{q}\sin(120^\circ-\theta)$ & 0  \\
atom 2 & $\frac{0.5}{q}\cos\theta$            & $\frac{0.5}{q}\sin\theta$            & 0 \\
atom 3 & $-\frac{0.5}{q}\cos(60^\circ-\theta)$  & $\frac{0.5}{q}\sin(60^\circ-\theta)$   & 0  \\   
\end{tabular}
\end{ruledtabular} 
\end{table} 

\begin{figure}
\includegraphics[width=5.0cm]{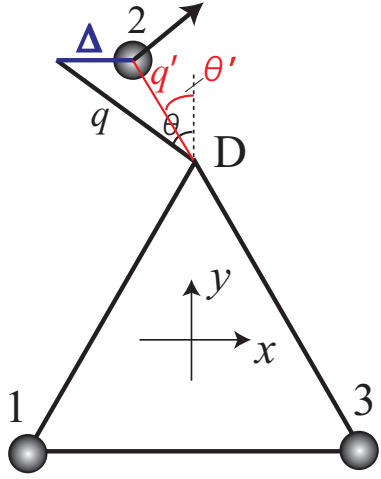}%
\caption{Schematic view of the derivation process of the $x$ component of second-order NAC on atom 2. 
A small displacement $\Delta$ is made in the $x$ direction for atom 2.  After displacement the contour 
radius is changed from $q$ to $q '$, and the contour angle is from $\theta$ to $\theta '$. 
The arrow denotes the new NAC vector on atom 2. 
\label{fig_JT_2nd_NAC_2x}
}
\end{figure}

To derive the $x$ component of second-order NAC on atom 2, we move atom 2 in the $x$ direction 
with a small displacement $\Delta$, as shown by Fig.~\ref{fig_JT_2nd_NAC_2x}. 
Since the Jahn-Teller model is a two-level system, we can get 
\begin{equation}
\left\langle \Psi_{0}\right\vert \frac{\partial ^2}{\partial x_{2}^2}\left\vert
\Psi_{1}\right\rangle = \frac{\partial}{\partial x_2}\left\langle \Psi_{0}\right\vert \frac{\partial}{\partial x_2}\left\vert
\Psi_{1}\right\rangle=\frac{A_{x_2}^\mathrm{disp}-A_{x_2}}{\Delta}, 
\label{eq_JT_x2}
\end{equation}
where $A_{x_2}$ is the $x$ component of first-order NAC before the 
displacement, as listed in Table~\ref{tab_JT_1st_NAC}.  
The new $x$ component of first-order NAC on atom 2 after the displacement, $A_{x_2}^\text{disp}$,   
can be determined from the new geometry as 
\begin{equation}
A_{x_2}^\text{disp}= \frac{0.5}{q '}\cos\theta ',   
\end{equation}
where  
\begin{equation}
q '=\sqrt{q^2\cos ^2 \theta +(q\sin\theta-\Delta)^2 }
\end{equation} 
and
\begin{equation}
\theta '=\arccos \left( \frac{q\cos\theta}{q '} \right).
\end{equation}
By taking $\Delta\to 0$ in Eq.~(\ref{eq_JT_x2}), we can get 
\begin{equation}
\left\langle \Psi_{0}\right\vert \frac{\partial ^2}{\partial x_{2}^2}\left\vert
\Psi_{1}\right\rangle =\lim_{\Delta\to 0}\frac{1}{\Delta}
\left(\frac{0.5}{q '}\cos\theta '
-\frac{0.5}{q}\cos\theta \right) 
=\frac{0.5}{q^2}\sin 2\theta
\end{equation}

\begin{figure}
\includegraphics[width=8.5cm]{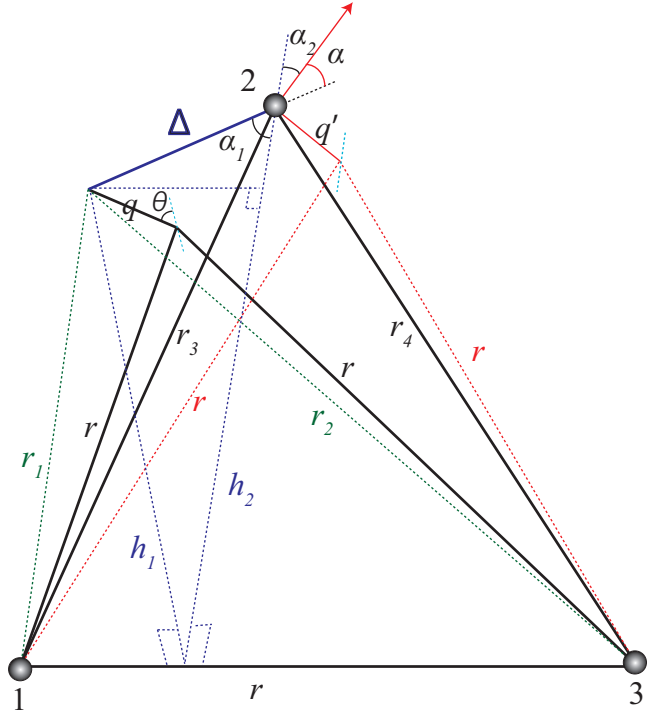}%
\caption{Schematic view of the derivation process of the $z$ component of 2nd-order NAC on atom 2. 
After a small displacement $\Delta$ is made in the $z$ direction for atom 2, the contour 
radius is changed from $q$ to $q '$, and the NAC vector (denoted by the arrow ending on atom 2) 
is located in the new atomic plane.    
\label{fig_JT_2nd_NAC_2z}
}
\end{figure}

The derivation of the $y$ component of second-order NAC on atom 2 is similar to that of the $x$ component 
in the above, thus the detail is not shown here. Next, to derive the $z$ component, we move atom 2 in the $z$ direction  
as shown by Fig.~\ref{fig_JT_2nd_NAC_2z}, and then we can get 
\begin{equation}
\left\langle \Psi_{0}\right\vert \frac{\partial ^2}{\partial z_{2}^2}\left\vert
\Psi_{1}\right\rangle = \frac{\partial}{\partial z_2}\left\langle \Psi_{0}\right\vert \frac{\partial}{\partial z_2}\left\vert
\Psi_{1}\right\rangle=\frac{A_{z_2}^\mathrm{disp}}{\Delta}, 
\label{eq_JT_z2}
\end{equation}
which uses the fact that the $z$ component of first-order NAC on atom 2 before the 
displacement is zero. 
The new geometry after the displacement gives 
\begin{equation}
A_{z_2}^\text{disp}= \frac{0.5}{q '}\cos\alpha=\frac{0.5}{q '}\cos\alpha _1 \cos\alpha _2,   
\end{equation}
where  $q '$ is the new contour radius around the vertex of the equilateral triangle in the new atomic plane.  
$\alpha$ is the angle between the new NAC vector and the $z$ axis, 
$\alpha _1$ is the angle between the new atomic plane and the $z$ axis, and 
$\alpha _2$ is the angle between the new NAC vector and the projection of the $z$ axis 
in the new atomic plane. Using the geometric relationships shown by Fig.~\ref{fig_JT_2nd_NAC_2z}, 
we can get 
\begin{equation}
q '=\sqrt{r_3^2+r^2-2r_3 r \cos\left[\arccos\left(\frac{r_3^2+r^2-r_4^2}{2r_3 r}\right) -60^\circ\right]}, 
\end{equation} 
\begin{equation}
\alpha _1=\arcsin\left(\frac{h_1}{h_2}\right), 
\end{equation}
and 
\begin{equation}
\alpha _2=90^\circ-\arccos\left(\frac{q'^2+r_4^2-r^2}{2q ' r_4}\right)-\arccos\left(\frac{h_2}{r_4}\right). 
\end{equation}
The auxiliary quantities in the above equations are calculated as 
\begin{equation*}
r_3=\sqrt{r_1^2+\Delta^2}, r_4=\sqrt{r_2^2+\Delta^2}, 
\end{equation*} 
\begin{equation*}
h_1=\frac{\sqrt{3}}{2}+q\cos\theta, h_2=\sqrt{h_1^2+\Delta^2},
\end{equation*}
\begin{equation*}
r_1=\sqrt{r^2+q^2-2qr\cos(150^\circ-\theta)},
\end{equation*}
\begin{equation*}
r_2=\sqrt{r^2+q^2-2qr\cos(210^\circ-\theta)}. 
\end{equation*} 
By taking $\Delta\to 0$, we can get 
$r_3 \to r_1$, $r_4 \to r_2$, $h_2 \to h_1$, 
 $q ' \to q$, and $\alpha _2 \to 90^\circ-\theta$. Then Eq.~(\ref{eq_JT_z2}) is 
reduced to  
\begin{equation}
\left\langle \Psi_{0}\right\vert \frac{\partial ^2}{\partial z_{2}^2}\left\vert
\Psi_{1}\right\rangle =\lim_{\Delta\to 0}\frac{1}{\Delta}
\frac{0.5}{q}\cos\left[\arcsin\left(\frac{h_1}{h_2}\right)\right]\cos(90^\circ-\theta) 
=\frac{0.5}{q}\frac{1}{\sqrt{3}r/2}\sin\theta, 
\end{equation}
where we have used the fact that $r \gg q$. 

\begin{figure}
\includegraphics[width=6.0cm]{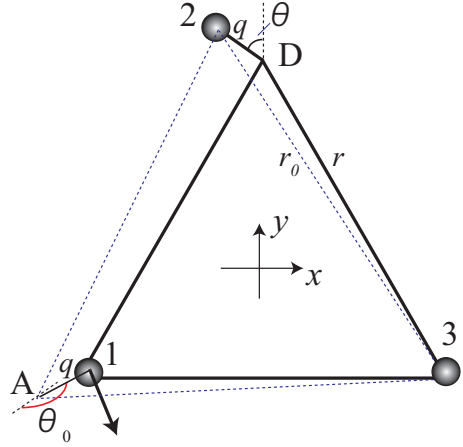}%
\caption{Schematic view of the derivation process of 2nd-order NAC components on atom 1, 
which is regarded as rotating around vertex A of a new equilateral triangle with side length $r_0$. 
The corresponding contour radius and angle is $q$ and  $\theta _0$, respectively. 
\label{fig_JT_2nd_NAC_1} 
}
\end{figure} 

To derive components of second-order NAC on atom 1 and atom 3, we need not 
make displacements but can merely use the fact that the three atoms are equivalent, i.e., 
not only atom 2 can be regarded as rotating in a contour around the intersection 
point, other two atoms can also be taken into such a view. In Fig.~\ref{fig_JT_2nd_NAC_1}, where atomic 
geometry is the same as in Fig.~\ref{fig_JT_1st_NAC}, atom 1 is regarded as rotating around the 
intersection point A with contour radius $q$ and angle $\theta _0$. 
Here A is the vertex of a new equilateral triangle with side length $r_0$. 
The geometric analysis gives 
\begin{equation}
r_0=\sqrt{q^2+r^2-2qr\cos(210^\circ-\theta)}, 
\end{equation}
and 
\begin{equation}
\theta_0=150^\circ+\arccos\left(\frac{q^2+r_0^2-r^2}{2qr_0}\right) 
=150^\circ+\arccos\left[\frac{q-r\cos(210^\circ-\theta)}{r}\right]. 
\end{equation}
Using the fact that $r \gg q$ we can easily get 
$r_0=r$ and $\theta_0=120^\circ+\theta$. 
Replacing $r$ and $\theta$ in the expression of second-order NAC components on atom 2 with 
$r_0$ and $\theta_0$, we can immediately get the results for atom 1. 

The results for atom 3 can be derived in a way similar to that for atom 1. 
The final results of second-order NAC components on all atoms are listed in Table~\ref{tab_JT_2nd_NAC}. 

\begin{table}
\caption{\label{tab_JT_2nd_NAC}The $x$, $y$ and $z$ components of second-order NAC on the 
three atoms of a Jahn-Teller trimer, which are in the geometry shown by Fig.~\ref{fig_JT_1st_NAC}.  
$q$ is the contour radius and $\theta$ is the contour angle.  
}
\begin{ruledtabular}
\begin{tabular}{lccc} 
 & $x$ & $y$ & $z$ \\
\hline
atom 1 & $\frac{0.5}{q^2}\cos(30^\circ+2\theta)$   & $-\frac{0.5}{q^2}\cos(30^\circ+2\theta)$ & $\frac{0.5}{q}\frac{1}{\sqrt{3}r/2}\sin(\theta+120^\circ)$  \\
atom 2 & $\frac{0.5}{q^2}\sin2\theta$              & $-\frac{0.5}{q^2}\sin2\theta$            & $\frac{0.5}{q}\frac{1}{\sqrt{3}r/2}\sin\theta$              \\
atom 3 & $-\frac{0.5}{q^2}\sin(120^\circ-2\theta)$ & $\frac{0.5}{q^2}\sin(120^\circ-2\theta)$ & $\frac{0.5}{q}\frac{1}{\sqrt{3}r/2}\sin(\theta-120^\circ)$  \\   
\end{tabular}
\end{ruledtabular} 
\end{table}

\section{\label{app_RT}Second-order NAC from the Renner-Teller model}

\begin{figure}
\includegraphics[width=8.5cm]{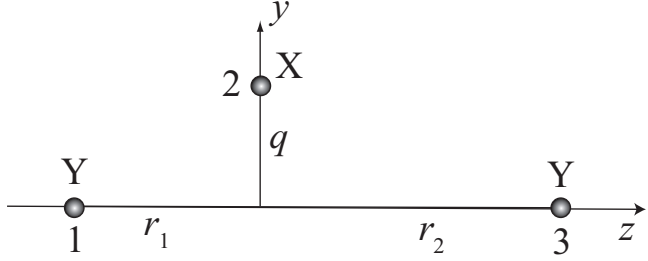}%
\caption{Configuration of an XY$_2$ Renner-Teller system in the $yz$ plane. The three atoms are located in a 
geometry slightly distorted from the linear geometry, and atom 2 is regarded on a 
contour around the $z$ axis with contour radius $q$.   
\label{fig_RT_1st_NAC}
}
\end{figure}

\begin{table}
\caption{\label{tab_RT_1st_NAC}The $x$, $y$ and $z$ components of first-order NAC on 
the three atoms of an XY$_2$ Renner-Teller system, which is in the geometry shown by Fig.~\ref{fig_RT_1st_NAC}.  
$q$ is the contour radius, while $r_1$ ($r_2$) is the distance of atom 1 (atom 3) 
from the intersection point. 
}
\begin{ruledtabular}
\begin{tabular}{lccc} 
 & $x$ & $y$ & $z$ \\
\hline
atom 1 & $-\frac{1}{q}\frac{r_2}{r_1+r_2}$  & 0  & 0  \\
atom 2 & $\frac{1}{q}$                      & 0  & 0 \\
atom 3 & $-\frac{1}{q}\frac{r_1}{r_1+r_2}$  & 0  & 0  \\   
\end{tabular}
\end{ruledtabular} 
\end{table} 

Figure~\ref{fig_RT_1st_NAC} shows an arbitrary configuration of an XY$_2$ Reller-Teller systems, where the X atom is 
moved with a sufficiently small displacement $q$ from the $z$ axis, which is the seam of Renner-Teller intersections.  
According to the Renner-Teller model \cite{lorquet}, the angular NAC has a quantized value of 1.0 
and all components of first-order NAC on three atoms can be determined, as shown by  
Table~\ref{tab_RT_1st_NAC}. Note that all $y$ and $z$ components are equal to zero, 
meaning that the NAC vectors are parallel to the $x$ axis.

Because the first-order NAC vectors are perpendicular to the $yz$ plane, small movement of atoms in the $yz$ plane 
will not alter the direction of NAC vectors and the $y$ and $z$ components of first-order NAC are kept to be zero after the displacement. 
Therefore, we can immediately conclude that $y$ and $z$ components of second-order NAC 
on all atoms are zero. 

\begin{figure}
\includegraphics[width=8.5cm]{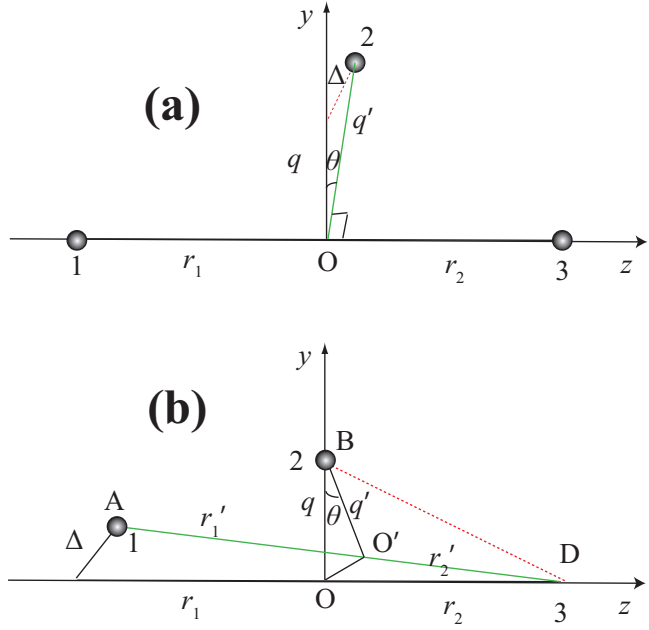}%
\caption{Schematic view of the derivation process of the $x$ component of second-order NAC on (a) atom 2 and (b) atom 1. 
A small displacement $\Delta$ is made in the $x$ direction for atom 2 in (a) and atom 1 in (b).  After displacement the contour 
radius is changed from $q$ to $q '$.  Note that in (b) the intersection point is changed from $\text{O}$ to $\text{O} '$.
\label{fig_RT_2nd_NAC_2x1x}
}
\end{figure}

To derive the $x$ component of second-order NAC on atom 2, we move atom 2 in the $x$ direction,  
as shown by Fig.~\ref{fig_RT_2nd_NAC_2x1x}(a). Since the Renner-Teller model is a two-level system, we can get  
\begin{equation}
\left\langle \Psi_{0}\right\vert \frac{\partial ^2}{\partial x_{2}^2}\left\vert
\Psi_{1}\right\rangle = \frac{\partial}{\partial x_2}\left\langle \Psi_{0}\right\vert \frac{\partial}{\partial x_2}\left\vert
\Psi_{1}\right\rangle=\frac{A_{x_2}^\mathrm{disp}-A_{x_2}}{\Delta}, 
\label{eq_RT_x2}
\end{equation}
where 
\begin{equation}
A_{x_2} = \frac{1}{q}.  
\end{equation}
The new geometry after the displacement gives 
\begin{equation}
A_{x_2}^\text{disp}= \frac{1}{q '}\cos\theta,   
\end{equation}
where  $q '=\sqrt{q^2 + \Delta ^2 } $ and $\theta=\arccos(q/q ')$. 
By taking $\Delta\to 0$ in Eq.~(\ref{eq_RT_x2}), we can get 
\begin{equation}
\left\langle \Psi_{0}\right\vert \frac{\partial ^2}{\partial x_{2}^2}\left\vert
\Psi_{1}\right\rangle =\lim_{\Delta\to 0}\frac{1}{\Delta}
\left[\frac{-\Delta ^2}{(q^2+\Delta ^2)q}\right]
=0.
\end{equation}

In a similar way the $x$ component of second-order NAC on atom 1 can be derived by displacing atom $1$ in the $x$ direction,  
shown by Fig.~\ref{fig_RT_2nd_NAC_2x1x}(b), as  
\begin{equation}
\left\langle \Psi_{0}\right\vert \frac{\partial ^2}{\partial x_{1}^2}\left\vert
\Psi_{1}\right\rangle = \frac{\partial}{\partial x_1}\left\langle \Psi_{0}\right\vert \frac{\partial}{\partial x_1}\left\vert
\Psi_{1}\right\rangle=\frac{A_{x_1}^\mathrm{disp}-A_{x_1}}{\Delta}, 
\label{eq_RT_x1}
\end{equation}
where 
\begin{equation}
A_{x_1} = -\frac{1}{q}\frac{r_2}{r_1+r_2},  
\end{equation}
\begin{equation}
A_{x_1}^\text{disp}= -\frac{1}{q '}\frac{r_2 '}{r_1 ' + r_2'}\cos\theta ,   
\end{equation} 
\begin{equation} 
q'=\sqrt{q^2+(r_2-r_2')^2}, 
\end{equation} 
\begin{equation}
\theta=\arccos(q/q'), 
\end{equation}
\begin{equation}
r_2'=r_2\cdot\frac{r_1+r_2}{r_1'+r_2'}=r_2\cdot\frac{r_1+r_2}{\sqrt{(r_1+r_2)^2+\Delta ^2}}. 
\end{equation} 
Taking $\Delta\to 0$ in Eq.~(\ref{eq_RT_x1}), we can get 
\begin{equation}
\left\langle \Psi_{0}\right\vert \frac{\partial ^2}{\partial x_{1}^2}\left\vert
\Psi_{1}\right\rangle =\lim_{\Delta\to 0}\frac{1}{\Delta}
\frac{q^2 r_2 \Delta ^2 + r_2^3 \Delta ^2}{\left[q^2 \Delta ^2 + q^2 (r_1+r_2)^2+r_2^2 \Delta ^2)\right]q(r_1+r_2)}
=0.
\end{equation}

Finally, since the derivation of $x$ component of second-order NAC on atom 3 is essentially 
equivalent to that of atom 1, the result for atom 3 is also 0.  

In one word, all components of second-order NAC on the three atoms of the Renner-Teller system 
are equal to zero.

\bibliography{second_NAC}


\end{document}